# Superconducting half-dome in bilayer nickelates

Yidi Liu[1,2*†], Bai Yang Wang[1,3*‡], Jiarui Li[1,3*§], Yaoju Tarn[1,3], Lopa Bhatt[4], Michael Colletta[4], Yi-Ming Wu[1,2], Sang-Jun Lee[5], Cheng-Tai Kuo[5], Jun-Sik Lee[5], Berit H. Goodge[6], David A. Muller[4,7], Zhi-Xun Shen[1,2,3,8], Srinivas Raghu[1,2,8], Harold Y. Hwang[1,3,8¶], and Yijun Yu[1,3‖]

[1]*Stanford Institute for Materials and Energy Sciences, SLAC National Accelerator Laboratory, Menlo Park, CA 94025, USA*

[2]*Department of Physics, Stanford University, Stanford, CA 94305, USA*

[3]*Department of Applied Physics, Stanford University, Stanford, CA 94305, USA*

[4]*School of Applied and Engineering Physics, Cornell University, Ithaca, New York 14850, USA*

[5]*Stanford Synchrotron Radiation Lightsource, SLAC National Accelerator Laboratory, Menlo Park, CA, USA*

[6]*Max Planck Institute for Chemical Physics of Solids, 01187 Dresden, Germany*

[7]*Kavli Institute at Cornell for Nanoscale Technology, Cornell University, Ithaca, New York 14850, USA*

[8]*Geballe Laboratory for Advanced Materials, Stanford University, Stanford, CA 94305, USA*

[*]These authors contributed equally to this work

[†]Contact author: yidil@stanford.edu

[‡]Contact author: bwang87@stanford.edu

[§]Contact author: jiaruili@stanford.edu

[¶]Contact author: hyhwang@stanford.edu

[‖]Contact author: yuyijun@stanford.edu

Understanding how superconductivity emerges and collapses in correlated electron systems remains a central challenge in condensed matter physics. As a recently discovered member of the high temperature superconductor family, bilayer nickelates provide a new opportunity for examining this problem. Their pronounced sensitivity to oxygen stoichiometry, while posing challenges for stabilizing superconductivity, simultaneously offers an effective control parameter for tuning electronic phases. Here we report a superconducting half-dome in compressively strained bilayer nickelate thin films as a function of continuous tuning of oxygen stoichiometry. Starting from an optimally superconducting state, increasing oxygen stoichiometry gradually suppresses superconductivity toward a metallic phase, whereas decreasing oxygen stoichiometry drives a granular superconductor-to-insulator transition while leaving the superconducting onset intact. This half-dome structure is consistent with the contrasting roles played by interstitial oxygen versus oxygen vacancies – namely the dominance of doping versus scattering. Notably, the half-dome emerges consistently across samples with different rare-earth combinations, with or without alkaline-earth doping, revealing a general feature of the bilayer nickelate phase diagram and informing opportunities for further enhancing superconductivity.

## I. INTRODUCTION

Establishing a generic phase diagram has been central in the study of high temperature superconductors across various materials families [1,2]. Such diagrams uncover the interplay between competing ground states, provide a unifying framework connecting disparate experimental observations, and guide the understanding of materials-related parameters. Comparing phase diagrams across different superconducting families further highlights universal features of high temperature superconductivity. Such an endeavor is now underway in the bilayer nickelates.

Since the discovery of superconductivity in bulk and thin film bilayer nickelates [3–7], their superconducting phase diagram has been explored along various dimensions including pressure [3–5,8–11], strain [6,7,12–15], oxygen stoichiometry [16–20], and substitutional hole doping [21]. For bulk samples, superconductivity is found to be broadly robust to variations in pressure above a threshold value, yet sensitive to oxygen off-stoichiometry [22–26] and crystal symmetry [27]. In thin films, a hole-doping-dependent superconducting dome is reported, with the optimal superconductivity transition temperature ($T_c$) observed at a finite doping level [21]. Superconductivity in thin films is also seen to positively correlate with compressive epitaxial strain [15] and pressure [14].

Bridging these disparate, partial phase diagrams and constructing a coherent view of superconductivity in bilayer nickelates is a high priority. Our work is motivated to unify aspects of the high-dimensional phase space of strain (or pressure), substitutional (Ca) hole doping, and oxygen stoichiometry. By examining thin film bilayer nickelates systematically across oxygen stoichiometry and substitutional doping, we find a universal superconducting half-dome, governed by oxygen stoichiometry in samples across different chemical doping levels, as summarized in Fig. 1.

Oxygen stoichiometry exhibits two qualitatively distinct manifestations in bilayer nickelates. Excess oxygen appears to predominantly act as an electronic tuning parameter, modifying carrier density via interstitial doping while gradually weakening superconductivity. In contrast, oxygen deficiency is associated with rapidly enhanced strong scattering and electronic inhomogeneity, suppressing coherent metallic transport while preserving local superconducting fluctuations, consistent with granular superconductivity [28]. This pronounced asymmetry suggests that oxygen vacancies play a fundamentally different role from oxygen interstitials, possibly reflecting the electronic fragility of the $NiO_2$ bilayer units, where apical and planar oxygen coordination is intimately tied to inter- and intra-layer coupling. In the following we present the experimental results that establish this phase diagram.

The starting point for this study is the realization that high crystallinity and oxygen stoichiometry are prerequisites for optimizing superconductivity in bilayer nickelate films [23]. Going beyond optimization, we see a trend in ozone-density [$w(O_3)$] versus annealing-temperature ($T_{anneal}$) that systematically varies $T_c$, as shown in Fig. 2(a). Notably, samples annealed closer to (but not beyond) the boundary between the bilayer phase and higher-order RP phases exhibit systematically reduced $T_c$, as reflected by the size of the filled circles (which is plotted proportional to $T_c$). We observe this trend in two different variations of bilayer nickelates, $La_2PrNi_2O_{7+\delta}$ (LPNO, Fig. 2(b)) and $La_2SmNi_2O_{7+\delta}$ (LSNO, Fig. 2(c)). We note that $\delta$ may take either positive or negative values throughout this manuscript corresponding to oxygen excess or deficiency, respectively. Importantly, the reduction in $T_c$ is not accompanied by a structural transformation into higher-order RP phases (Fig. S1 [29]). This provides strong evidence that superconductivity is also influenced by the evolution of oxygen stoichiometry within the stable bilayer phase (a similar trend also observed in ref. [20]).

## II. RESULTS

### A. Excess oxygen in metallic bilayer nickelate thin films

While $La_2NiO_{4+\delta}$ is known to accommodate interstitial excess oxygen [30], initial studies of bulk bilayer nickelates reported a clear tendency toward apical oxygen vacancies [22,31]. Given that the bilayer nickelate structure incorporates both perovskite and rock-salt building blocks, it can host either oxygen deficiency or oxygen excess. Indeed, recent studies show evidence of interstitial oxygen for both bulk and thin film bilayer nickelates [24–26,32]. Thus, we deduce that the suppression of superconductivity in samples undergoing more oxidizing ozone annealing conditions in Fig. 2 arises from excess oxygen intercalation.

Directly probing oxygen content in thin films is challenging, and currently impossible at the required accuracy for these ultrathin films whose thickness is limited to the regime of coherent strain. Unlike bulk cuprates [33] or nickelates [10,11], where thermogravimetric analysis or chemical titration can accurately determine the oxygen content, only a limited set of techniques can be applied here; we use spectroscopy to probe the evolution of the oxygen stoichiometry by comparison to bulk data.

First, we establish a correspondence between an optimally superconducting LPNO film (identified through transport) and its oxygen stoichiometry. Specifically, 'optimally superconducting' refers to samples with $T_{c,onset}$ > 40 K and a narrow transition width (i.e. $T_{c,90\%}$ - $T_{c,10\%}$ < 10 K; the various criteria used to define $T_c$ throughout the manuscript are illustrated in Fig. S2 [29]) just after ozone annealing. Electron energy-loss spectroscopy (EELS) was performed on such an LPNO sample for comparison to bulk reference data correlating the O *K*-edge pre-peak intensity with oxygen stoichiometry [22,25] (Fig. 3(a)). The specimen was prepared by cryogenic focused ion beam (cryo-FIB) to eliminate processing-induced modification of the oxygen stoichiometry (see Methods), and spectra were spatially averaged over the bottom half of the film to exclude measurements of possible oxygen loss near the surface. The measured pre-peak intensity aligns with the $\delta \approx 0$ reference bulk sample (Fig. 3(a)), indicating that optimally superconducting samples are (spectroscopically) very similar to the stoichiometric composition $La_2PrNi_2O_7$. By analogy, we infer that stoichiometric $La_2SmNi_2O_7$ samples also possess optimal superconducting properties, given that the optimal superconducting LSNO films have similar metrics of $T_c$ and Hall coefficient ($R_H$) to those of the LPNO counterparts.

Next, we examine the X-ray absorption spectra (XAS) of samples subjected to different oxygen-treatment conditions, which show different transport behaviors. In Fig. 3(b)-(c), we show transport data of an insulating sample (purple thick line; oxygen-deficient), an optimally superconducting sample (green thick line; stoichiometric, as argued above), a non-superconducting metallic sample (red thick line; after ozone annealing near the phase boundary in Fig. 2(a), oxygen-rich), and four samples in between (see Appendix for details). Ni $L_2$-edge XAS evolves from the $Ni^{2+}$-like spectrum in the oxygen-deficient sample to a more $Ni^{3+}$-like spectrum in the oxygen-rich sample (Fig. 3(d)), consistent with expectations from formal electron counting. This trend is further illustrated by O *K*-edge XAS spectra in Fig. 3(e), where the oxygen pre-peak intensity increases from the insulating sample to the optimally superconducting sample and finally to the non-superconducting metallic sample, indicating a systematic increase in oxygen content. Using the oxygen pre-peak intensity as a spectroscopic proxy for oxygen stoichiometry, we plot the corresponding transport data as a contour plot in Fig. 3(f). Although coarsely sampled, it already exhibits a half-dome evolution as we will discuss further in section B below. Such changes in oxygen stoichiometry are also accompanied by subtle structural changes, including a slight difference in the $c$-lattice constant. However, conflicting reports of the $c$-dependence on oxygen stoichiometry exist in the literature for both bulk and thin films. A detailed discussion is in Fig. S3 [29].

Before closing this section, several observations are noteworthy: (i) Similar to bulk results [11], LSNO films consistently exhibit better crystallinity than LPNO. (ii) This, in turn, allows LSNO films to approach the phase boundary more closely without transforming into mixed higher-order RP phases and, as a result, these specimens intercalate more oxygen and exhibit a lower $T_c$ as shown in Fig. 2(b)-(c). In the extreme cases, superconductivity disappears. (iii) A lower residual resistance ratio (RRR, defined as $\rho$(200 K)/$\rho$(60 K)) is observed in samples with lower $T_c$, suggesting slightly increased scattering.

### B. Evolution of superconductivity with oxygen stoichiometry in LSNO

Having benchmarked oxygen stoichiometry for discrete samples spanning the endpoints of metallic and insulating behavior, we examine the evolution of a ***single*** sample under continuous (and reversible, see Fig. S4 [29]) tuning of oxygen stoichiometry between these endpoints. Beginning from the state with the highest excess oxygen, stepwise vacuum annealing (see Appendix) reveals a two-stage evolution of transport behavior. In the first stage (Stage I, Fig. 4(a)), the gradual removal of oxygen induces superconductivity from a non-superconducting metallic ground state. Further removal of oxygen (Stage II, Fig. 4(b)) rapidly increases resistivity and eventually drives the system through a superconductor-insulator transition (SIT) [16,17,19], while the resistivity drop at $T_{c,onset}$ remains robust. By aligning the $\rho(T)$ curves of the three LSNO samples plotted in thick lines in Fig. 3(b)-(c) (red, green, and purple) with those of the same sample measured at the beginning and end of Stage I and II in Fig. 4(a)-(b) (also red, green, and purple, respectively), a clear correspondence emerges: the stoichiometric point delineates the crossover between the two stages.

Plotting the conductivity at 200 K, $\sigma$(200 K), against the cumulative annealing time $t_{anneal}$ (Fig. 4(d); the $x$-axis is reversed so positive $\delta$ appears on the right) further highlights a distinct crossover separating the two stages: $\sigma$(200 K) decreases linearly with $t_{anneal}$ in Stage I, but switches to a more rapid, exponential decay in Stage II. Since $\Delta\delta(t_{anneal}) \sim \exp(-\lambda t_{anneal})$ under surface-exchange-limited first-order kinetics [34,35], $t_{anneal}$ serves as a smooth, monotonic proxy for $\delta$ (to leading order $t_{anneal} \sim -\Delta\delta$). Therefore, the two distinct stages observed in the transport evolution and $\sigma$(200 K) with $t_{anneal}$ can also be directly mapped to the evolution with $\delta$. Interestingly, the intersection of the two fitting functions for each stage in Fig. 4(d) coincides with the maximum RRR as a function of $t_{anneal}$ shown in Fig. 4(c), offering a further indication of the mechanism governing the crossover.

A superconducting half-dome is clearly visible in the resistivity contour (Fig. 4(f)) and in the corresponding second derivative contour (Fig. 5(d)), where the $x$- and $y$-axis represent reversed $t_{anneal}$ (as a proxy for $\delta$) and temperature, respectively. The extracted $T_{c,onset}$ and $T_{c,50\%}$ are overlaid on the contours. We use the term "half-dome" (further justified in Fig. 5) to indicate that on the oxygen-rich side both $T_{c,onset}$ and $T_{c,50\%}$ decrease in tandem, whereas on the oxygen-deficient side $T_{c,onset}$ remains nearly unchanged while $T_{c,50\%}$ collapses approaching the insulating state. The measurement with the highest $T_{c,10\%}$ (as a proxy for global phase coherence) across all annealing steps is marked as 'optimized'.

The similarity of the half-dome structure in Fig. 4(f) and Fig. 3(f) further motivates exploration of the possibility of using an experimentally accessible quantity directly related to oxygen stoichiometry, as a more intrinsic axis for the phase diagram. To this end, the O $K$-edge XAS pre-peak intensity (Fig. 3(e)) serves as a sensitive proxy for oxygen stoichiometry as in cuprates [36–38]. By comparing the transport evolution of the seven coarsely sampled points presented in Fig. 3(b)-(c) (replotted in Fig. S5(a) [29]) with the finely sampled annealing-dependent transport data presented in Fig. 4(a)-(b) (replotted in Fig. S5(b) [29]), we establish the correspondence between $t_{anneal}$ and the O $K$-edge pre-peak intensity (Fig. S5(c) [29]), allowing the sample-dependent annealing parameter to be converted into a more physically meaningful oxygen-related parameter (Fig. 4(f), the blue-shaded scale).

To further estimate the oxygen stoichiometry, we calibrated the pre-peak intensity using two reference points: the optimally superconducting sample, which is close to oxygen stoichiometry, and the metallic sample, whose oxygen stoichiometry is estimated based on comparison with Ca-doped samples, which will be discussed in Sec. II D. The remaining points were interpolated or extrapolated by these two references (see caption of Fig. 4). This two-point calibration provides a reasonable and physically motivated estimate of the oxygen stoichiometry in this challenging thin-film system. Based on this calibration, we are able to further map the phase diagram as a function of the estimated oxygen stoichiometry (Fig. 4(f), the brown-shaded scale).

We now discuss the structure of this superconducting phase diagram in more detail below. In Stage I ($\delta > 0$), as $\delta$ increases from stoichiometric ($\delta \approx 0$), the right sector of the superconducting half-dome evolves in a manner reminiscent of overdoped cuprates [39], where both $T_{c,onset}$ and $T_{c,50\%}$ decrease, shifting in parallel toward zero temperature while maintaining a similar transition width (from green to red curves in Fig. 4(a), see also Fig. 4(f)). When the excess oxygens occupy the interstitial sites between the rock-salt layers [24,25,32], their primary effect is consistent with doping the adjacent $NiO_2$ bilayers through a 'modulation doping' mechanism—analogous to doping by interstitial oxygen in BSCCO [39,40]. This is supported by the Drude-like linear increase of $\sigma$(200 K) with increasing $\delta$ (Fig. 4(d)). There is in addition (weak) disorder, evidenced by the slight decrease in RRR with increasing $\delta$ (Fig. 4(c), also mentioned earlier when discussing Fig. 2(b)-(c)).

In Stage II ($\delta < 0$), as $\delta$ decreases from near stoichiometric ($\delta \approx 0$), apical and/or planar oxygens are removed, as observed in refs. [21,22,16], which is expected to disrupt the structural continuity of the $NiO_2$ bilayer as an integral building block. In cuprates, oxygen vacancies within $CuO_2$ planes are also known to be highly detrimental to superconductivity, primarily through the introduction of strong disorder [41]. Similar vacancy formation in nickelates is therefore expected to produce comparable deleterious effects. Moreover, vacancies at the apical oxygen sites—which are essential for maintaining interlayer coupling within the bilayer structure—have been proposed to be catastrophic for superconductivity [42,43]. Consistently, we see a rapid increase of resistivity (from green to purple curves in Fig. 4(b)), an exponential decrease of $\sigma$(200 K) (Fig. 4(d)), and a fast decrease in RRR (Fig. 4(c)). While $T_{c,onset}$ remains constant or even slightly increases, $T_{c,50\%}$ abruptly drops to zero (Fig. 4(f)), indicative of the onset of granular superconductivity [28]. This ultimately leads to an SIT: finite-size scaling analysis yields the critical exponents product $z\nu = 2.5$, consistent with the quantum percolation model and previous reports [16] (Fig. S6 [29]).

In the 'Crossover' region between Stages I and II ($\delta \approx 0$), $T_c$ remains nearly constant, with a slight increase as oxygen is removed. Due to entropy considerations and potential spatial inhomogeneity, oxygen vacancies likely start forming before all interstitial oxygens are completely removed. The peak in RRR can be naturally interpreted as a balance point where disorder from both oxygen vacancies and interstitial oxygen is minimized; we therefore label this moment as 'balanced'. Notably, in Fig. 4(f), the $t_{anneal}$ for the 'balanced' point lies slightly on the oxygen-rich side of the 'optimized' point. This offset can be understood by considering that oxygen vacancies introduce much stronger disorder, such that their influence outweighs the weaker effects of residual interstitial oxygen, effectively shifting the 'balanced' point slightly toward the oxygen-rich side.

Taken together, these three regimes distinguish the oxygen-tuned superconducting phase diagram of bilayer nickelates from the canonical superconducting dome of cuprate from ref. [44]. Figure 5 makes this distinction explicit through a side-by-side comparison. In cuprates, $T_{c,onset}$ and $T_{c,50\%}$ evolve together throughout the evolution of the oxygen stoichiometry (Fig. 5(a)). Thus, the negative regions (red) of the second derivative contour of $\rho$ (Fig. 5(c)) form a closed dome. In bilayer nickelates, by contrast, $T_{c,onset}$ remains robust while $T_{c,50\%}$ collapses (Fig. 5(b)) during the removal of oxygen. The corresponding red regions of the second derivative contour of $\rho$ remain open and gradually fade toward the oxygen-deficient side (Fig. 5(d)). This behavior is consistent with the persistence of local superconducting fluctuations while global coherence is progressively suppressed. Thus, the half-dome reflects contrasting evolution between a predominantly doping-like suppression on the oxygen-excess side and coherence-limited granular superconductivity on the oxygen-deficient side, rather than merely an incomplete contour.

Additional insight can be found in the evolution of $R_H$ with $t_{anneal}$ or change of $\delta$ (Fig. 4(e)). Positive [6,45], near-zero [7,21], and negative [20,23] $R_H$ values have been observed in superconducting bilayer nickelate thin films. A recent study also explored the evolution of oxygen-deficient regions [17]. Here, by precisely tuning $\delta$ from excess, to stoichiometric, and finally to a deficient regime, we reproduce all previously reported trends in $R_H$—spanning negative to near-zero to positive (see Fig. S7 for detailed comparison with literature [29]). This complete evolution enables us to pinpoint the negative $R_H$ (Fig. S7(b), red thick line [29]) corresponding to the optimally superconducting (stoichiometric) sample, as also observed in refs. [20,23]. Another notable feature on the $\delta > 0$ side is that $R_H$ reaches its negative minimum and then increases, coinciding with the point when reduction of $T_c$ occurs (but not with the moment of maximum RRR). As discussed below, this minimum is observed during the change of $\delta$ in bilayer nickelate films with different substitutional doping, which provides a way to quantify the effect of $\delta$.

### C. Oxygen stoichiometry tuning in the presence of calcium doping

Cuprates often exhibit similar phase diagrams upon carrier doping, whether achieved through alkaline-earth substitution or interstitial oxygen intercalation, as exemplified by $La_{2-x}Sr_xCuO_4$ and $La_2CuO_{4+\delta}$ (ref. [46]). An intriguing question is whether this correspondence also holds for bilayer nickelates, and the potential dual contributions to doping from both cation substitution and oxygen stoichiometry [47]. To address this, we performed extensive investigations of the oxygen-stoichiometry-dependent transport properties of $La_{2-x}PrCa_xNi_2O_{7+\delta}$ (LPCNO) with different Ca doping levels ($x$ = 0, 0.165, 0.382, $\rho(T)$ shown in Fig. S8 [29]) and compared them with those of LSNO.

Resistivity contour plots were constructed as in Fig. 4 for each composition (Fig. 6). To enable a meaningful comparison, however, $t_{anneal}$ is not a suitable variable, as both the oxygen loss rate and absolute resistivity are composition and sample dependent. To address this, two normalization procedures were applied. (i) To compensate for differences in oxygen loss rates among samples, $\sigma$(200 K) was used as a monotonic proxy for $t_{anneal}$ (Fig. 4(d)), and thus for $\delta$. (ii) To correct for variations in absolute resistivity, previous studies on cuprates with varying oxygen stoichiometry renormalized $\sigma$(200 K) with respect to the optimally doped state, which serves as a natural reference point [40,48]. In the present case, however, such a well-defined anchor is absent. Instead, we found that the occurrence of an $R_H$ minimum during vacuum annealing is a robust and generic feature across samples with different compositions or deposition methods. Therefore, $\sigma$(200 K) was normalized to its value at the $R_H$ minimum. This procedure provides a reproducible reference frame for comparing the evolution of transport properties across samples.

Now we present the contour map of resistivity, normalized to its value at 200 K, for the same LSNO sample from Fig. 4, and three LPCNO samples at different doping levels (Fig. 6(a)-(d)) and the corresponding evolutions of $R_H$ (Fig. 6(e)–(h)). To frame a more general discussion of the superconducting half-domes, we highlight four key observations in Fig. 6. (i) The superconducting region consistently exhibits a half-dome-like structure as discussed earlier, indicating a similar underlying mechanism across changes in the oxygen stoichiometry. (ii) The superconducting half-domes are less complete in LPCNO than those of LSNO. This is likely because ozone annealing near the phase boundary is less robust for LPCNO, as discussed previously in Fig. 2. Consequently, we reduced the scale of annealing compared to that for LSNO, which in turn led to a narrower range of resistivity tuning via changing $\delta$. We also note that the smaller ion size of Sm than Pr likely allows for a higher concentration of interstitial oxygens. (iii) The right sector of the half-dome becomes more complete with increasing Ca doping, which is consistent with higher initial hole doping. (iv) Although determining the precise stoichiometric point is challenging in this system, we can use the 'optimized' (as well as the 'balanced') criteria introduced in Fig. 4, as an ansatz for the stoichiometric composition. If this assumption holds, the corresponding 'optimized' ('balanced') point on $\sigma(200\ K)_{norm}$ axis should shift toward higher values with increasing Ca doping, which is indeed what is observed.

### D. General superconducting half-dome in bilayer nickelates

The above observations imply that the interstitial oxygen plays a role analogous to Ca doping, as increasing either one drives the system towards the right sector of the superconducting half-dome. Given that a doping level of 0.19 holes per Ni in the LPCNO sample almost fully suppresses superconductivity, we infer that $\delta > 0.19$ in the non-superconducting metallic LSNO film, assuming each interstitial oxygen contributes two holes. This is roughly consistent with the bulk value of $\delta = 0.23 \pm 0.07$, where superconductivity is completely suppressed even under pressure [25]. Beyond this carrier-doping perspective, recent bulk studies have also shown that interstitial oxygen can distort the Ni–O bond angle, potentially suppressing superconductivity [24,25]. A similar mechanism may occur in our films; however, unlike in the bulk, where interstitial oxygen stabilizes an orthorhombic phase even under high pressure, epitaxial strain confines films to the tetragonal phase, with the associated interstitial oxygen occupancy configuration yet to be determined.

Returning to Fig. 1, a general phase diagram of samples with different compositions is plotted against $\sigma(200\ K)_{norm}$ and temperature. $T_{c,onset}$ and the metal-insulator transition temperature ($T_{ins}$, defined as the temperature at which $\rho(T)$ reaches its minimum in the normal state; see Fig. S8 [29]) are displayed for all samples with different compositions. For $\delta > 0$, $T_{c,onset}$ of all different compositions collapses onto a common curve, forming the right sector of a superconducting half-dome. This indicates that this part of the half-dome represents a quasi-universal behavior of bilayer nickelates, irrespective of Ca doping. As discussed above, the primary effect of Ca doping is to shift the stoichiometric point rightward in the phase diagram shown in Fig. 1. The collapse of the data also supports the validity of the $\sigma$(200 K) normalization procedure, as it self-consistently indicates that the observed trend change in the Hall channel reflects an underlying generic behavior. We speculate that this feature may correspond either to a Lifshitz transition associated with the Fermi level crossing of the $\gamma$ band upon hole doping, or to the onset of a structural transition upon oxygen intercalation [25]. The former interpretation appears broadly consistent with the scenario proposed in ref. [49], although the precise correspondence remains to be clarified. On the left side, when $\delta < 0$, an SIT emerges, but the critical point varies from sample to sample, reflecting differences in the disorder scale. At fixed $\sigma(200\ \mathrm{K})_{\mathrm{norm}}$, samples with higher Ca doping show higher $T_{\mathrm{ins}}$ and lower RRR, both suggesting that Ca substitution introduces additional disorder (Fig. S8 [29]). While the occurrence of the SIT is a general feature, the critical $\sigma(200\ \mathrm{K})_{\mathrm{norm}}$ that triggers the SIT is thus sample-dependent.

## III. DISCUSSION

Overall, the robust half-dome-like evolution of $T_c$ points to an intrinsic asymmetry in the underlying physics, consistent with a scenario in which oxygen interstitials predominantly act as dopants, whereas oxygen vacancies introduce strong disorder [50]. This pronounced sensitivity of superconductivity to vacancies may be related to the important roles of the inner apical and planar oxygen sites in mediating strong Ni–O hybridization and interlayer coupling within the bilayer. Within this picture, their removal simultaneously reconstructs the relevant $e_g$ orbital and generates strong local disorder, both of which could be highly detrimental to superconducting coherence [42,43].

As for the phase competing with superconductivity in this regime, a natural candidate is a spin-density-wave (SDW) state [51–53]. Oxygen vacancies can play two roles: they decrease the effective Ni valence and introduce quenched randomness along the in-plane Ni–O network. These effects suggest at least two possible routes toward understanding the observed granularity of superconductivity. First, neglecting quenched disorder, oxygen vacancies act as an electron dopant. Recent results show the ground state electronic configuration is sensitive to oxygen vacancies [54]. Filling the apical ligand hole via electron doping effectively modulates the $d_{z2}$-$2p_z$-$d_{z2}$ interlayer exchange coupling. Moreover, the reduced Ni valence favors $Ni^{2+}$ higher-spin states [55], which are susceptible to translational-symmetry-breaking order with insulating ground states over a large range of doping [56], thereby suppressing superconductivity locally. Second, local oxygen vacancies create a local defect that breaks interlayer coupling, reducing the electronic dimensionality, which enhances electron correlations. When quenched randomness and bond disorder are included, the granularity is further enhanced by a rounding of the first-order transition between superconductivity and SDW order with disorder-induced coexistence of both phases [57]. Moreover, quenched disorder drives the metallic SDW electrons toward Anderson localization, weakening Josephson coupling between spatially separated superconducting regions [58,59]. In both pictures, superconductivity can only nucleate in isolated regions where magnetic order is locally suppressed, for example near magnetic domain walls [60,61]. Global superconducting coherence would then arise from Josephson tunneling through metallic SDW regions separating superconducting puddles.

More broadly, despite the distinct global superconducting phase diagrams of cuprates and bilayer nickelates discussed above, this phenomenology bears a conceptual resemblance to an earlier line of thought in the cuprate literature [62], where the underdoped regime has been discussed in terms of spatially inhomogeneous or granular superconductivity [63]. Although such a granular picture is not generally regarded as a universal description of the underdoped cuprates today, the phenomenological analogy is noteworthy: superconducting pairing may persist locally even as global phase coherence is limited by inhomogeneity and scattering. The similarity of this framework to the oxygen-deficient regime of bilayer nickelates points to a possible commonality in how disorder, reduced carrier mobility, and phase fluctuations shape superconducting behavior in oxides, without implying identical phase diagrams or microscopic origins in the two material families. Beyond nickelates and cuprates, analogous behavior has been reported across diverse superconducting systems, including Pb–Ag alloys [28], $(Ba,K)BiO_3$ (ref. [64]), $Li_xZrNCl$ (ref. [65]), and (Li,Fe)OHFeSe (ref. [66]). This suggests that asymmetric superconducting evolution driven by the interplay of doping, disorder, and competing phases is a broader organizing principle across superconductors. A final noteworthy point concerns the prospects of further raising $T_c$. The continued rise of $T_{c,\mathrm{onset}}$ toward the insulating regime on the left side of the phase diagram (Fig. 5(d)) indicates that the $T_c$ has not yet peaked within the doping window accessible by oxygen tuning, suggesting that the optimal composition lies beyond it. Instead of closing into the full dome familiar from the cuprates, the superconducting region terminates abruptly, defining only the overdoped half of a dome (Fig. 5). We attribute this to the onset of oxygen-vacancy disorder before a homogeneous, optimally doped state can be stabilized: disorder pre-empts, rather than forbids, access to a higher-$T_c$ homogeneous phase. Consistent with this picture, we do not observe the strange-metal transport reported near optimal doping in the cuprates and in bulk bilayer nickelates with higher $T_c$ under high pressure [5]. An even higher $T_c$ may therefore be attainable if electron doping can be introduced into bilayer nickelates without significant disorder. In this context, the synthesis of nickelates with new structural motifs represents a promising pathway toward this objective [67].

## APPENDIX: METHODS

### 1. Sample preparation and XRD characterization

The $SrLaAlO_4$(001) [SLAO(001)] substrates (MTI Corporation) were sonicated in acetone and IPA for 1 minute each prior to both PLD and MBE growth. For PLD growth, polycrystalline LSNO and LPNO were used as targets during the deposition process. The LSNO and LPNO films were deposited by PLD with a KrF excimer laser (wavelength 248 nm). During the growth, the substrate temperature was maintained at 660 °C under oxygen partial pressure $p(O_2)$ of 150 mTorr. The pulsed laser repetition rate during the deposition was 5 Hz. The laser fluence was 0.56 J/cm$^2$ with a spot size of 3.3 mm$^2$. For MBE growth, the SLAO substrate was first annealed in $5\times10^{-6}$ Torr ozone at 750 °C for 10 minutes. LPNO and LPCNO thin films were then deposited layer by layer at 680 °C in $1.8\times10^{-6}$ Torr ozone in a shutter-controlled mode. During growth, reflection high-energy electron diffraction with a 15 keV electron beam was employed to monitor the growth process and film quality. Unlike previous work [6,23], here we do not use a surface capping layer to facilitate oxygen tuning. For both PLD and MBE grown samples, XRD data were measured using a monochromated Cu $K\alpha1$ source ($\lambda$ = 1.5406 Å) after growth.

### 2. XAS measurements

XAS experiments were carried out at Beamline 13-2 at $T$ = 80 K and Beamline 13-3 at $T$ = 13.5 K of the SSRL. At both beamlines, absorption spectra were acquired in total electron-yield mode, with the $\sigma$-polarized X-rays incident at 20-degree grazing angle. The seven measurements shown in Fig. 3 were performed on four physical pieces cut from the same parent sample, thereby minimizing variations in initial sample quality. To accommodate the limited beamtime, the measurements were conducted in two batches. In the first batch, the four pieces, prepared with different oxygen stoichiometries by ozone annealing, were measured. The oxygen stoichiometry of three pieces was then modified by further vacuum annealing, and these pieces were remeasured in the second batch. To preserve oxygen stoichiometry, we adopted the procedures described in [23] during sample storage and transportation.

### 3. Transport measurements

Au electrodes (40 nm thick) were deposited using electron beam evaporation and then electrodes were bonded to ceramic chip carriers with aluminum wires by an ultrasonic wire bonder before performing ozone annealing. During ozone annealing, electrical transport was measured inside a tube furnace [23]. Low-temperature magnetotransport measurements were performed using $^4$He cryostats. The vacuum annealing was then performed within the same $^4$He cryostats by raising the temperature to ~360 K and holding at that temperature ($T_{anneal}$) before subsequent low-temperature measurements.

### 4. EELS measurements

To prepare cross-sectional scanning transmission electron microscopy (STEM) specimens, we used an FEI Strata 400S focused ion beam (FIB) system with a Quorum PP3010T cryo-stage. The FIB specimen stage was cooled to -192 °C while performing a cryogenic lift-out protocol adapted from ref. [68]. EELS of the O $K$-edge was measured on the cryo-FIB prepared lamella using a FEI Titan Themis operated at 300 kV accelerating voltage and 21.4 mrad semi-convergence angle equipped with a Gatan Quantum 965 energy filter and a K2 Summit direct electron detector. A Gatan Model 626 cryo-transfer holder was used to maintain specimen temperature below -180 °C throughout the measurement. Keeping the sample cool during the FIB process and STEM-EELS measurements preserves oxygen stoichiometry which is altered if performed at room temperature. Spectra were recorded with an energy dispersion of 0.1 eV per channel, 0.7 Å spatial step size, 3710 energy channels, and a 75 pA probe current.

## ACKNOWLEDGEMENTS

We thank W. He, S. A. Kivelson, Y. S. Lee, H. Luo, R. Peng, H. C. Xu, and Y. Zhang for discussions and assistance. This work was supported by the US Department of Energy, Office of Basic Energy Sciences, Division of Materials Sciences and Engineering (contract no. DE-AC0276SF00515), as well as the Kavli Foundation, Klaus Tschira Stiftung, and Kevin Wells (development of in-situ annealing). Work at the Stanford Nano Shared Facilities (SNSF) RRID:SCR_023230 is supported by the National Science Foundation under grant ECCS-1542152. B.H.G. was supported by the Max Planck Society. L.B. and D.A.M. acknowledge support by the NSF

Platform for the Accelerated Realization, Analysis, and Discovery of Interface Materials (PARADIM) under cooperative agreement No. DMR-2039380. STEM related work made use of the Cornell Center for Materials Research shared instrumentation facility. The FEI Titan Themis 300 was acquired through NSF-MRI-1429155, with additional support from Cornell University, the Weill Institute, and the Kavli Institute at Cornell. Support for the FIB/SEM cryo-stage and transfer system was provided by the Kavli Institute at Cornell and the Energy Materials Center at Cornell, DOE EFRC BES (DE-SC0001086). X-ray measurements were carried out at the SSRL, SLAC National Accelerator Laboratory, supported by the US Department of Energy, Office of Science, Office of Basic Energy Sciences (contract no. DE-AC02-76SF00515).

## AUTHOR CONTRIBUTIONS

Y.L., B.Y.W., and J.L. contributed equally to this work. Y.Y., H.Y.H. and Z.-X.S. supervised this project. B.Y.W., Y.L., J.L., H.Y.H., and Y.Y. conceived and designed the study. Y.L. synthesized LSNO and LPNO films, and B.Y.W. synthesized LPNO and LPCNO films. Y.Y., Y.L. and Y.T. studied the ozone annealing effects. Y.Y., Y.L., and B.Y.W. characterized and studied the transport properties with the assistance of Y.T. L.B., M.C., B.H.G. and D.A.M. performed EELS studies. J.L. performed the XAS measurements with support from S.J.L., C.T.K. and J.S.L. Y.L., B.Y.W., J.L., Y.-M.W., S.R., H.Y.H. and Y.Y. analyzed the data. Y.L., B.Y.W., J.L., S.R., H.Y.H. and Y.Y. wrote the paper with input from all authors.

## DATA AVAILABILITY

The data presented in the figures and other findings of this study are available from the corresponding authors upon reasonable request.

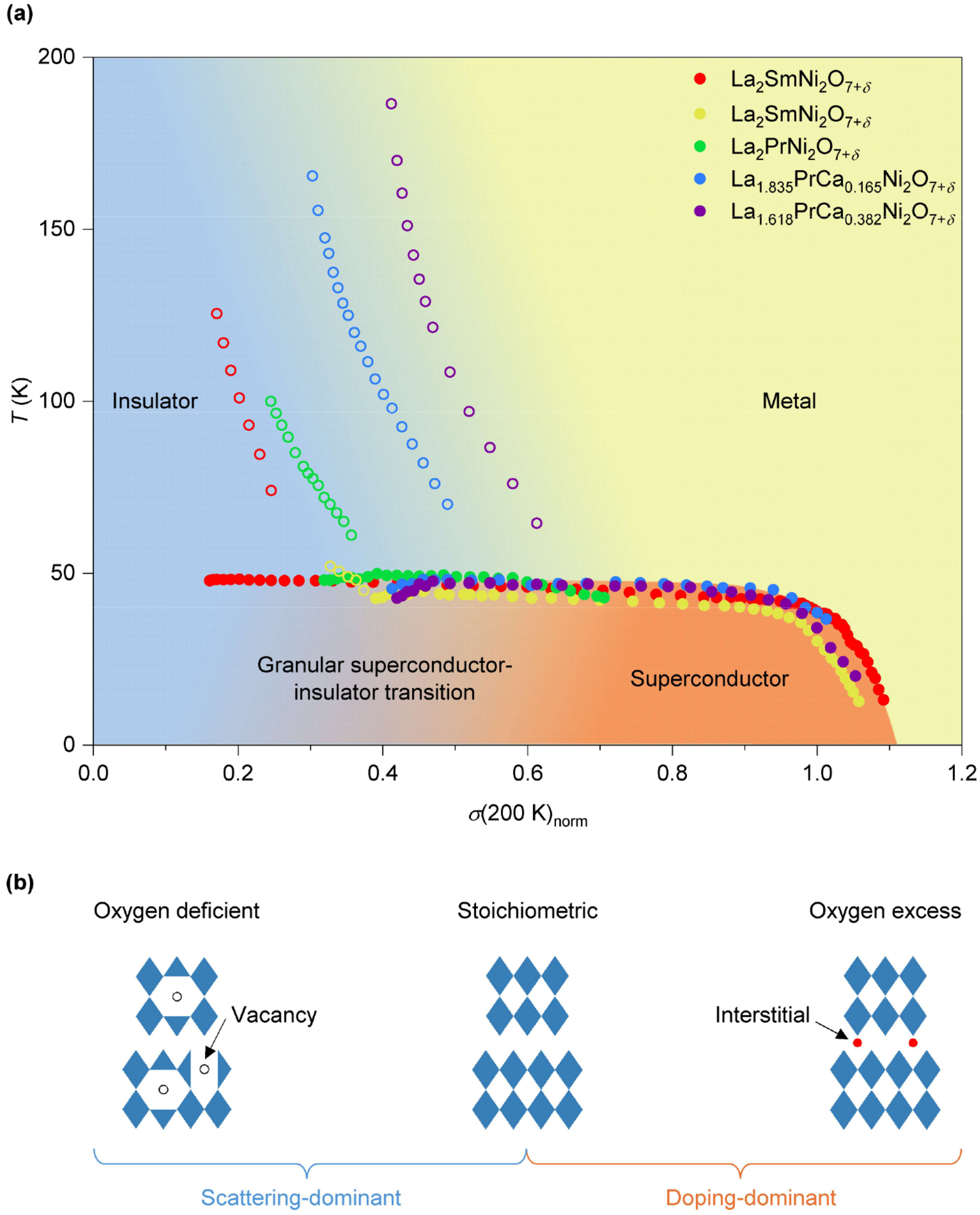

FIG. 1. Superconducting phase diagram of bilayer nickelate films. (a) General phase diagram plotted against $T$ and $\sigma(200\ \mathrm{K})_{\mathrm{norm}}$ obtained from samples studied in this work. Here, $\sigma(200\ \mathrm{K})_{\mathrm{norm}}$, defined in the main text, serves as a proxy for oxygen stoichiometry, with oxygen content increasing from left (deficient) to right (excess). $T_{\mathrm{c,onset}}$ and $T_{\mathrm{ins}}$ are labeled with filled and open circles. As the oxygen content increases from the deficient regime toward stoichiometry, a granular superconducting phase emerges from the insulator (blue), followed by the establishment of a robust superconducting state (red). Further oxygen incorporation beyond stoichiometry suppresses superconductivity, and the ground state evolves to a metal (yellow). Notably, granular superconductivity emerges exclusively on the oxygen-deficient side, giving rise to a half-dome-like structure. (b) Schematic of the structural evolution under varying oxygen stoichiometry. The contrasting roles of scattering versus doping are indicated. On the oxygen-deficient side, both apical and planar oxygen vacancies may be present [21,22,16]; therefore, both types are included in the schematic.

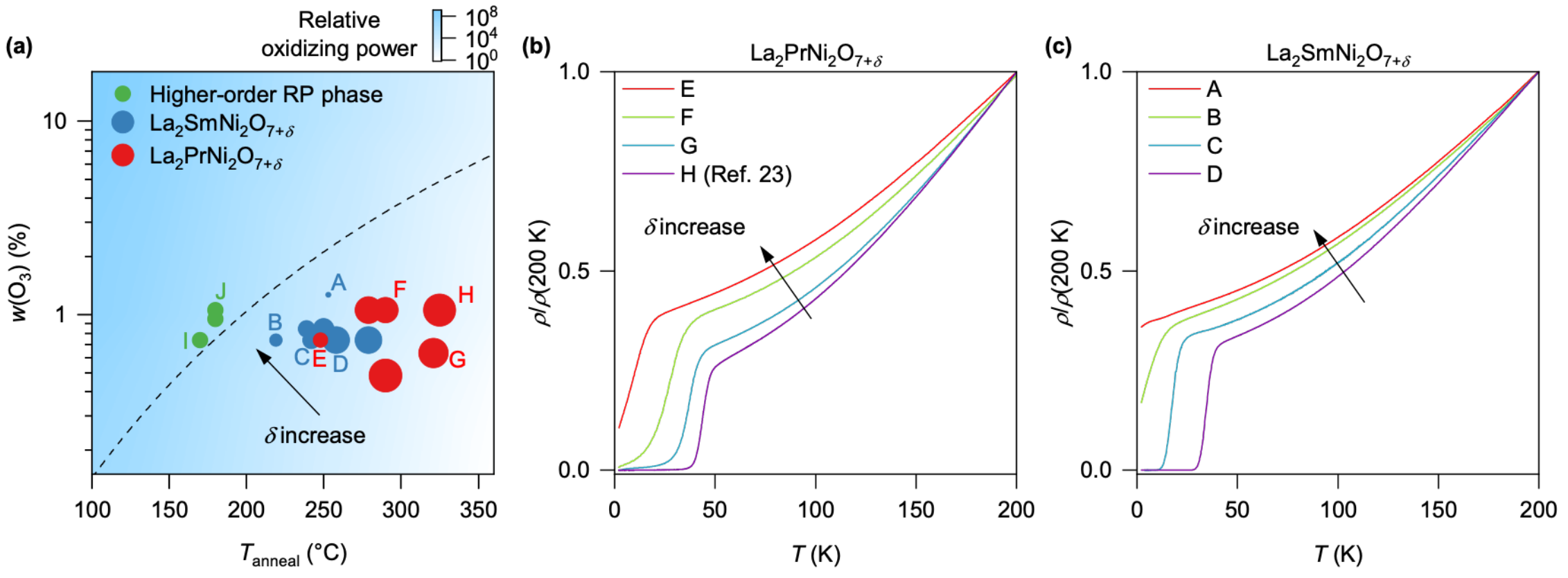

FIG. 2. Superconductivity of bilayer nickelate thin films under different ozone annealing conditions. (a) Phase diagram of ozone annealing temperature $T_{\mathrm{anneal}}$ vs. ozone concentration $w(O_3)$. Filled green circles represent films transformed into higher-order RP phases, while blue and red dots correspond to LSNO and LPNO films remaining in the bilayer phase. The size of each blue and red dot reflects the $T_{\mathrm{c,onset}}$ of each sample. The black dashed line represents the estimated stability boundary from ref. [23], above which the bilayer phase cannot be maintained; approaching this boundary allows more oxygen insertion. (b)-(c) $\rho(T)$ normalized at 200 K for LPNO and LSNO under different ozone conditions, respectively. Both $T_{\mathrm{c,onset}}$ and the normal-state resistivity evolve systematically with annealing conditions. Each curve corresponds to a sample labeled with the same letter as in (a).

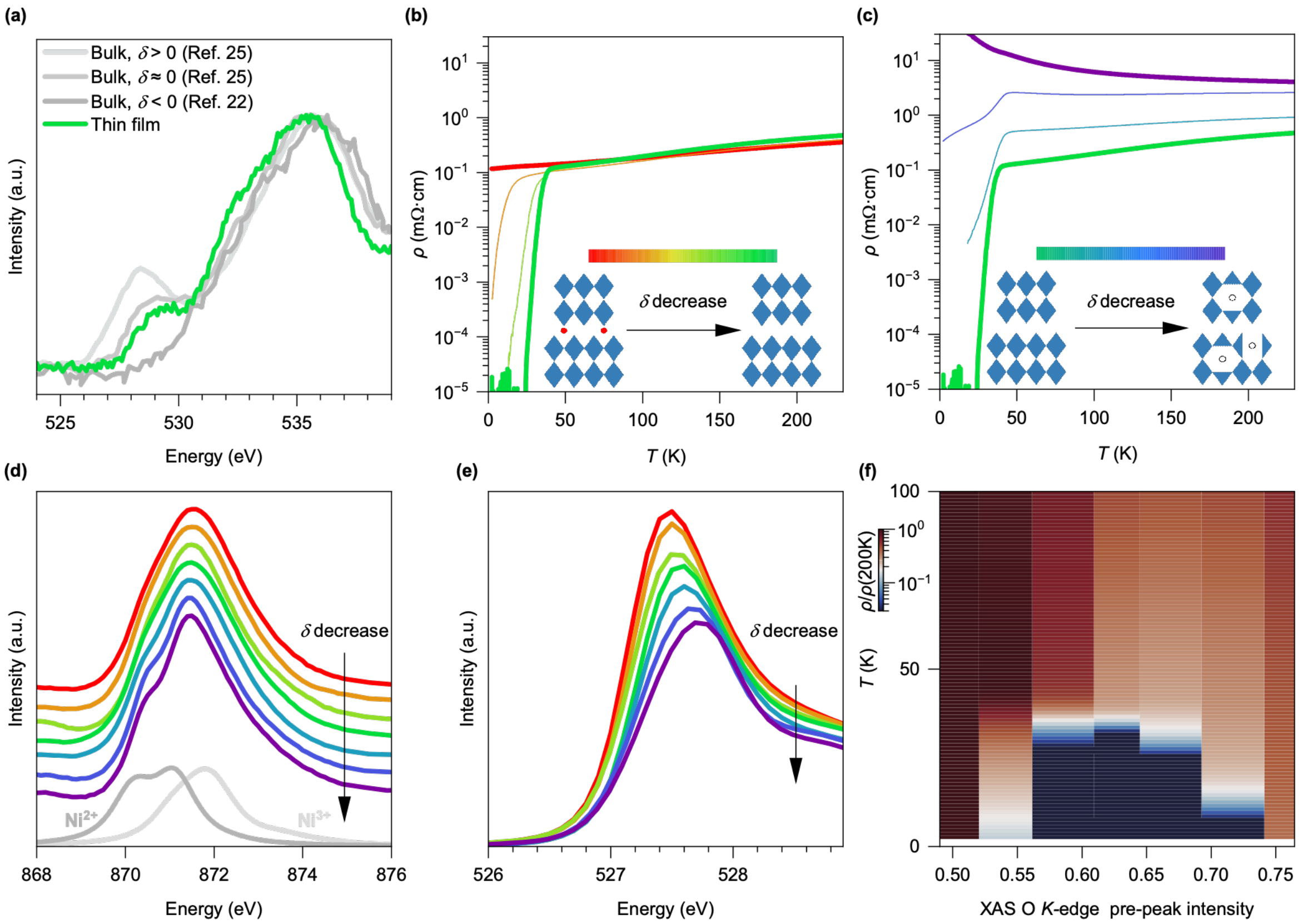

FIG. 3. Spectroscopic benchmarking of oxygen stoichiometry in bilayer nickelate thin films. (a) O *K*-edge EELS spectra of LPNO thin film sample with optimal superconductivity. Light grey, grey, and dark grey curves are reference spectra of bulk bilayer nickelates with positive, near zero, and negative $\delta$ from refs. [22,25]. (b), (c) $\rho(T)$ curves of LSNO thin films under seven different oxygen-treatment conditions, with colors denoting different oxygen stoichiometries. (b) shows the evolution from the metallic state after ozone annealing to the optimal superconducting state, while (c) shows the subsequent evolution from the optimal superconducting state to the insulating state. The inset schematics in (b) and (c) illustrate the structural evolution accompanying changes in oxygen stoichiometry. (d) Ni $L_2$-edge XAS spectra of the same samples in (b), (c). Dark and light grey lines are reference spectra for $Ni^{2+}$ and $Ni^{3+}$ from refs. [53,69]. (e) O *K*-edge XAS spectra of the same samples in (b), (c). (f) Contour plot of $\rho/\rho$(200 K) plotted as a function of the XAS O *K*-edge pre-peak intensity and *T*. The map exhibits features qualitatively similar to those observed in Fig. 4(f), enabling a comparison between the two phase diagrams.

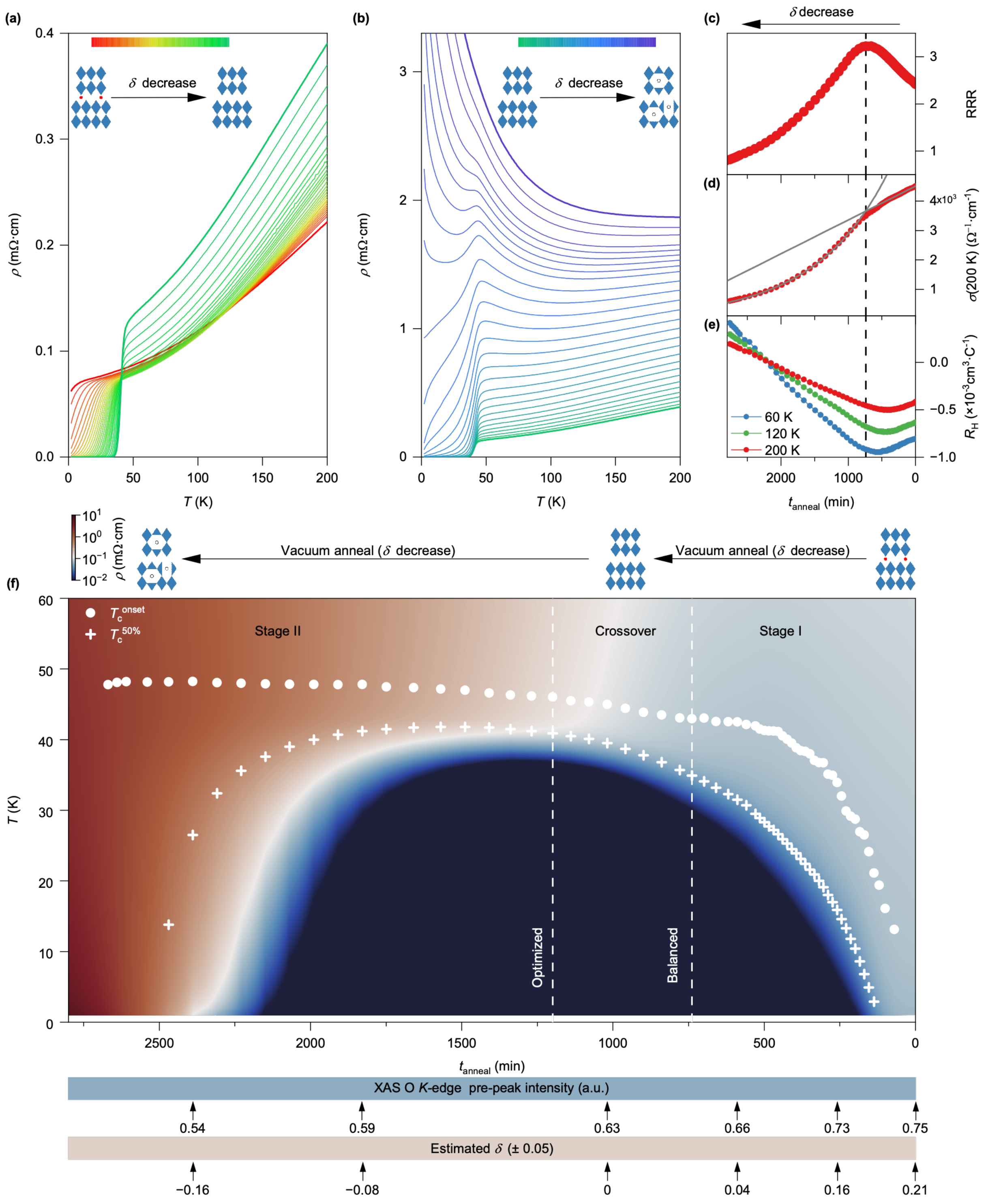

FIG. 4. A superconducting half-dome tuned by oxygen stoichiometry in an LSNO thin film. (a)-(b) $\rho(T)$ of an ozone-annealed LSNO thin film sample measured after sequential vacuum annealing. (a) shows the evolution from the metallic state to optimized superconductivity (Stage I and 'Crossover'), while (b) shows further evolution from optimized superconductivity to the insulating state ('Crossover' and Stage II). (c)-(d) Evolution of RRR and $\sigma$(200 K) plotted against $t_{\mathrm{anneal}}$, respectively. In (d), the grey curves indicate the fitted results using different functionals corresponding to Stages I and II for $\sigma$(200 K), respectively. The vertical dashed line aligns the intersection of two fittings in (d) and coincides with the maximum of RRR in (c). (e) Evolution of $R_{\mathrm{H}}$ at 60, 120, and 200 K plotted against $t_{anneal}$. (f) Contour map of $\rho$ plotted against $T$ and $t_{anneal}$, showing a superconducting half-dome. The inset schematics in (a), (b), and (f) sketch the structural evolution under oxygen stoichiometry change. The dashed lines indicate the positions for 'optimized' and 'balanced' states, with the 'Crossover' region defined between them. Below the $t_{anneal}$ axis, two additional scales are shown: the XAS O $K$-edge pre-peak intensity mapped to selected $t_{anneal}$ points (blue-shaded scale, estimated from Fig. S5 [29]) and the estimated $\delta$ (brown-shaded scale). $\delta$ was calibrated using two reference points: i) the optimally superconducting sample is anchored to $\delta \approx 0$ (Fig. 3); ii) the most oxygen-excess metallic sample is anchored to $\delta \approx 0.21 \pm 0.05$. The latter estimate is based on two considerations. First, our Ca-doped samples demonstrate that superconductivity persists up to a hole concentration of approximately 0.19 holes/Ni, indicating that the metallic state requires a hole doping level exceeding this value. Second, bulk LPNO references appear metallic for $\delta$ around 0.23 [25]. Therefore, the metallic state is expected to emerge within the range $\delta \approx$ 0.19–0.23. Here, we use the relation holes/Ni $\approx \delta$, considering that each interstitial oxygen contributes two holes and that there are two Ni ions per formula unit.

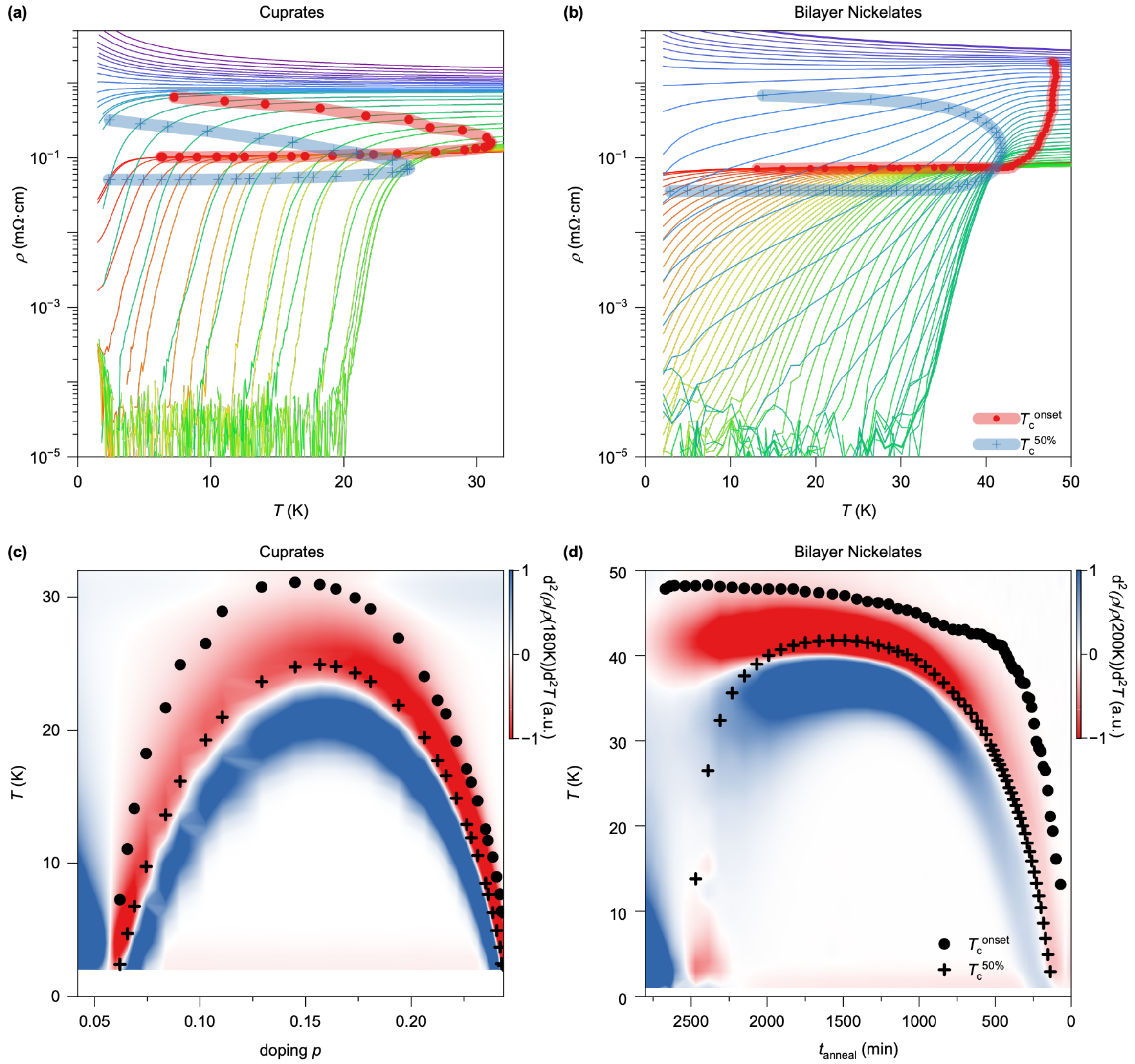

FIG. 5. Superconducting half-dome in bilayer nickelates compared with the superconducting dome in cuprates. Evolution of $\rho(T)$ for (a) cuprates, replotted from ref. [44] and (b) bilayer nickelates, replotted from Fig. 4(a)-(b), respectively. Red and blue bands show the evolution of $T_{c,\mathrm{onset}}$ and $T_{c,50\%}$, respectively. In cuprates, both $T_{c,\mathrm{onset}}$ and $T_{c,50\%}$ follow the same trend, whereas in bilayer nickelates they evolve in tandem on the oxygen-excess side but diverge on the oxygen-deficient side—a behavior we term a "half-dome". (c) Contour map of the second derivative of $\rho/\rho$(180 K) as a function of $T$ and doping, showing a full superconducting dome for the cuprate data in (a). (d) Contour map of the second derivative of $\rho/\rho$(200 K) as a function of $T$ and $t_{\mathrm{anneal}}$, showing a superconducting half-dome for the sample in (b).

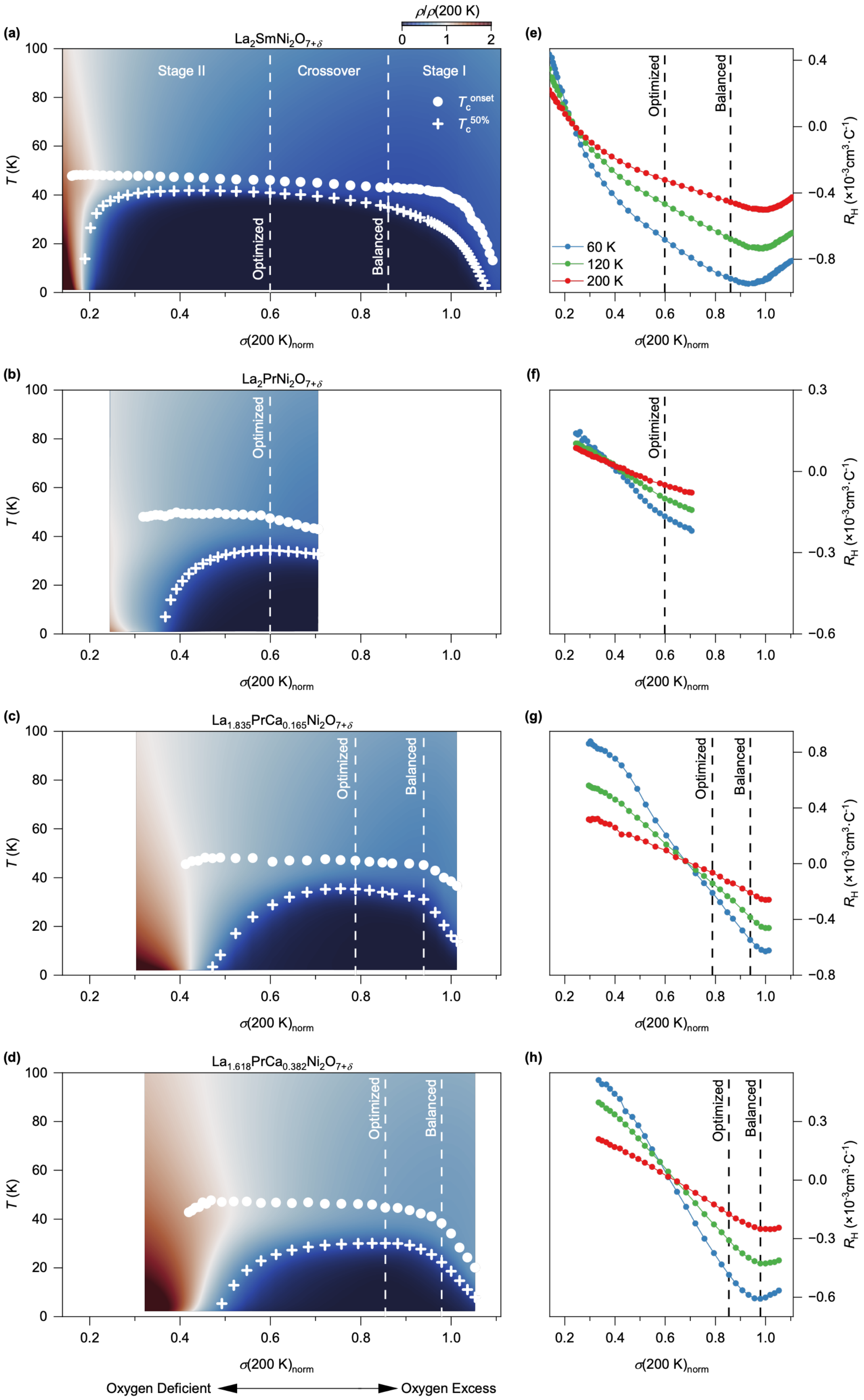

FIG. 6. Superconducting half-domes in LSNO and LPCNO films. (a) Contour maps of $\rho/\rho$(200 K) plotted against $T$

and $\sigma$(200 K)$_{norm}$ of the same sample in Fig. 4. (b)-(d) Contour maps of $\rho/\rho$(200 K) plotted against $T$ and $\sigma$(200 K)$_{norm}$ of LPCNO thin films with $x$ = 0, 0.165, 0.382, respectively. Each shows a similar but incomplete half-dome-like contour as in (a). (e)-(h) Evolution of $R_H$ at 60, 120, and 200 K plotted against $\sigma$(200 K)$_{norm}$ for the same samples in (a)-(d), respectively. All exhibit similar trends, with slight variations in the magnitude of $R_H$ among samples. The dashed lines indicate the positions of 'optimized' and 'balanced' states for each sample. $\sigma$(200 K) in (b) is scaled such that the point corresponding to optimized superconductivity aligns with that of LSNO in (a), as no local minimum is observed in $R_H$. Since no well-defined maximum in RRR is observed in (b), the 'balanced' position is not labeled.

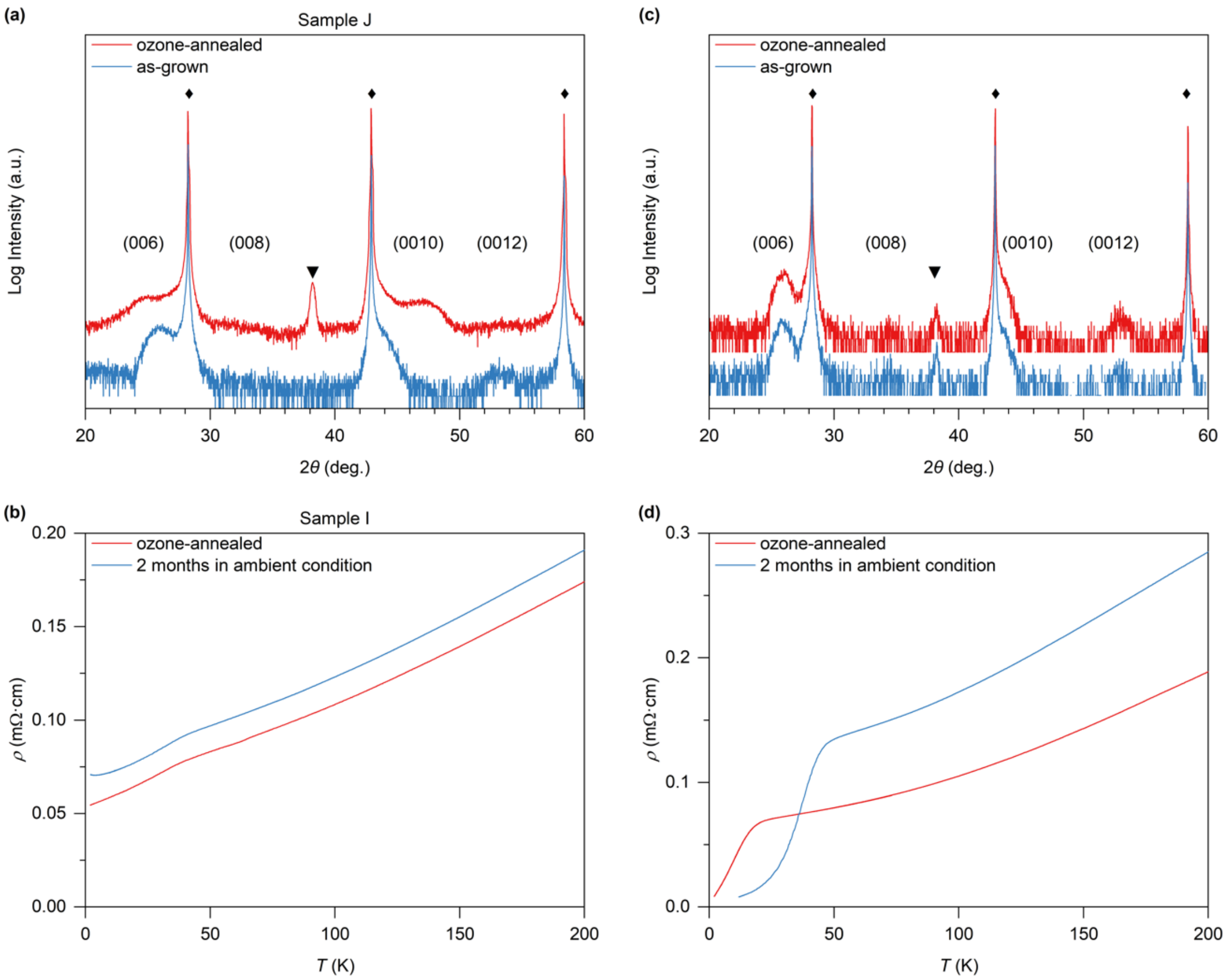

FIG. S1. Transport and structural evolution of bilayer nickelate thin films under different ozone annealing conditions. (a), (c) XRD $\theta$-$2\theta$ symmetric scans of LPNO thin film samples before and after ozone annealing. In (a), aggressive ozone annealing triggers the transformation to mixed higher-order RP phases, while in (c), the bilayer structure is preserved but with excess oxygen. The SLAO(001) substrate peaks are marked by ♦. The gold (111) peak is marked by ▼. (b), (d) $\rho(T)$ of LPNO thin film samples measured after ozone annealing and subsequently after two months in ambient conditions. If the material is (partially) transformed into higher-order RP phases (a), superconductivity remains absent after oxygen loss (b). In contrast, for samples in which the bilayer structure is preserved (c), $T_c$ is enhanced upon oxygen loss (d). Data plotted in (a),(b) are from the same samples labeled I and J in Fig. 2(a).

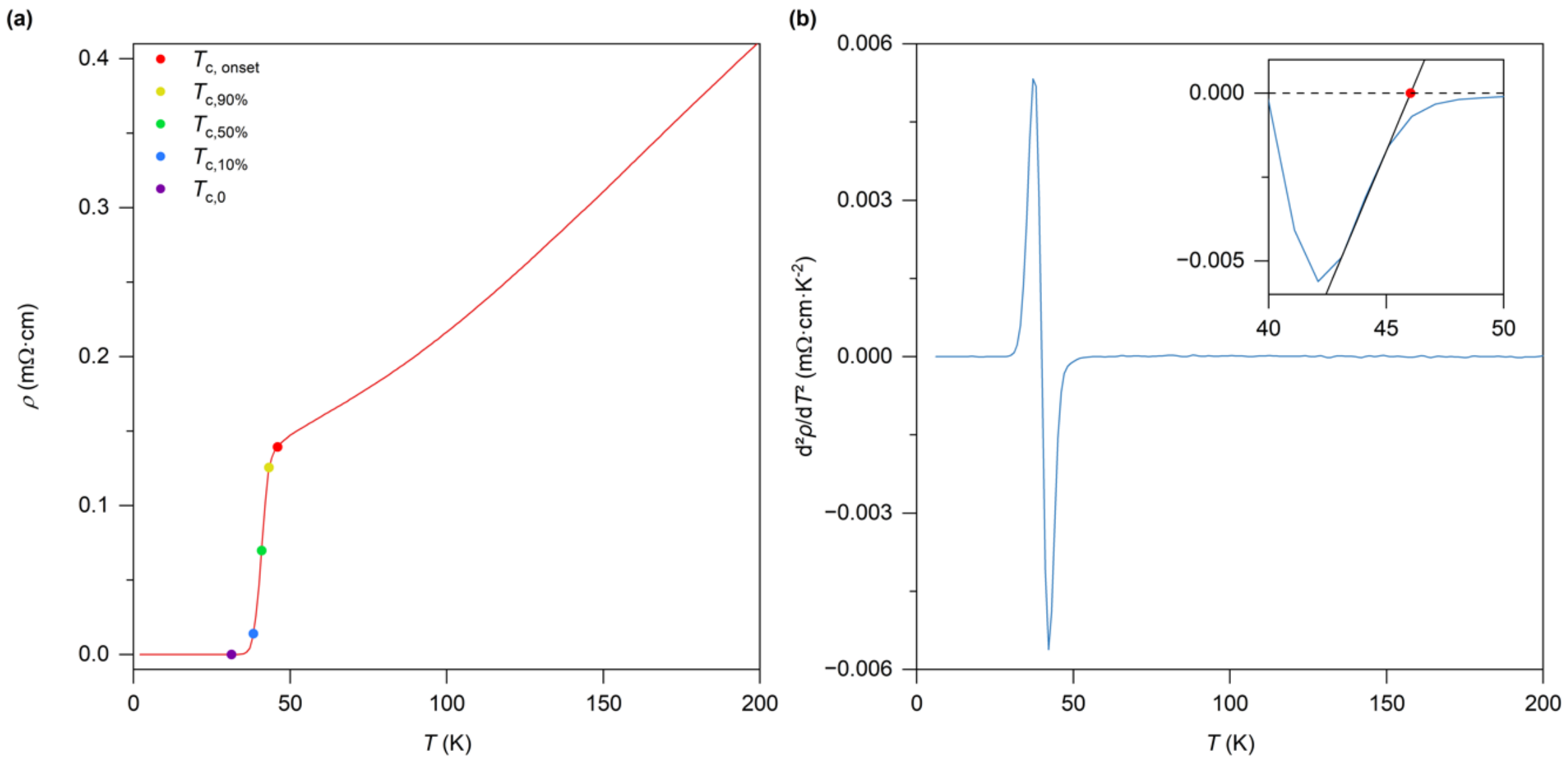

FIG. S2. Definition of transition temperatures. (a) $\rho(T)$ of an optimally superconducting LSNO thin film sample. The red, yellow, green, blue, and purple filled circles represent $T_{c,onset}$, $T_{c,90\%}$, $T_{c,50\%}$, $T_{c,10\%}$, and $T_{c,0}$, respectively. $T_{c,90\%}$, $T_{c,50\%}$, and $T_{c,10\%}$ are defined as the temperatures where the resistance reaches 90%, 50%, and 10% of the normal-state resistance near $T_{c,onset}$. (b) $d^2\rho/dT^2$ of the sample in (a). Inset: $T_{c,onset}$ is obtained by linearly fitting to the approximately linear region near the onset of the transition, with the zero intercept of the fit defining $T_{c,onset}$.

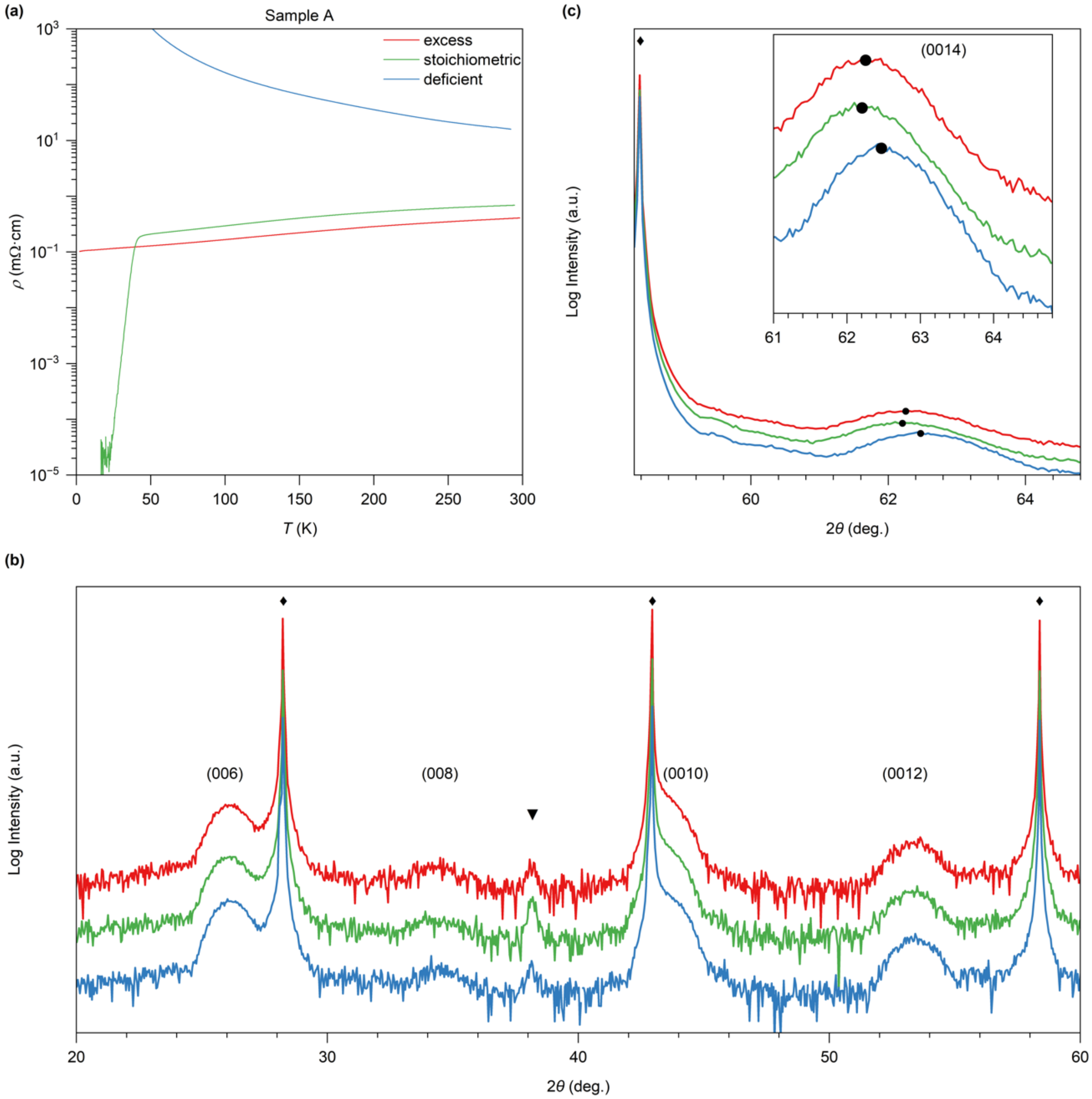

FIG. S3. Evolution of XRD patterns of bilayer nickelate thin films with different oxygen stoichiometries. (a) $\rho(T)$ of one LSNO thin film sample (labeled as sample A in Fig. 2(a)), measured after sequential post-growth treatments: as-grown (labeled ‘deficient’), after ozone annealing to a metallic state (labeled ‘excess’), and after vacuum-annealing to optimal superconductivity (labeled ‘stoichiometric’). (b) XRD $\theta$-$2\theta$ symmetric scans of LSNO sample after each step of post-growth treatments. The SLAO(001) substrate peaks are marked by ♦, and the gold (111) peak is marked by ▼. (c) XRD $\theta$-$2\theta$ symmetric scans of the sample over a narrower $2\theta$ range. The inset zooms in on the (00$\underline{14}$) film peak, marked by ●. A $c$ lattice expansion is observed as the samples evolve from an oxygen-deficient insulating state to a stoichiometric optimally superconducting state, consistent with ref. [18]. Subsequently, a very slight contraction is observed upon transitioning to the oxygen-excess metallic state, consistent with recent bulk reports [18,25]. However, the evolution of $c$ lattice with oxygen stoichiometry is not consistent across different literature studies. For instance, contrary to our observation, a bulk report suggested that slightly oxygen-deficient samples possess a larger $c$ lattice than stoichiometric ones [31]. Additionally, another report on thin films has shown a pronounced $c$ lattice expansion, rather than contraction, upon intercalating interstitial oxygen [32]. These discrepancies likely stem from multiple factors. The presence of polymorphism and mixed RP phases in early studies hinders exact stoichiometry determination in bulk, and variations in structural symmetry with oxygen content complicate lattice analysis [26,32]. Moreover, the epitaxial strain in thin films likely induces different structural responses to oxygen stoichiometry compared to bulk.

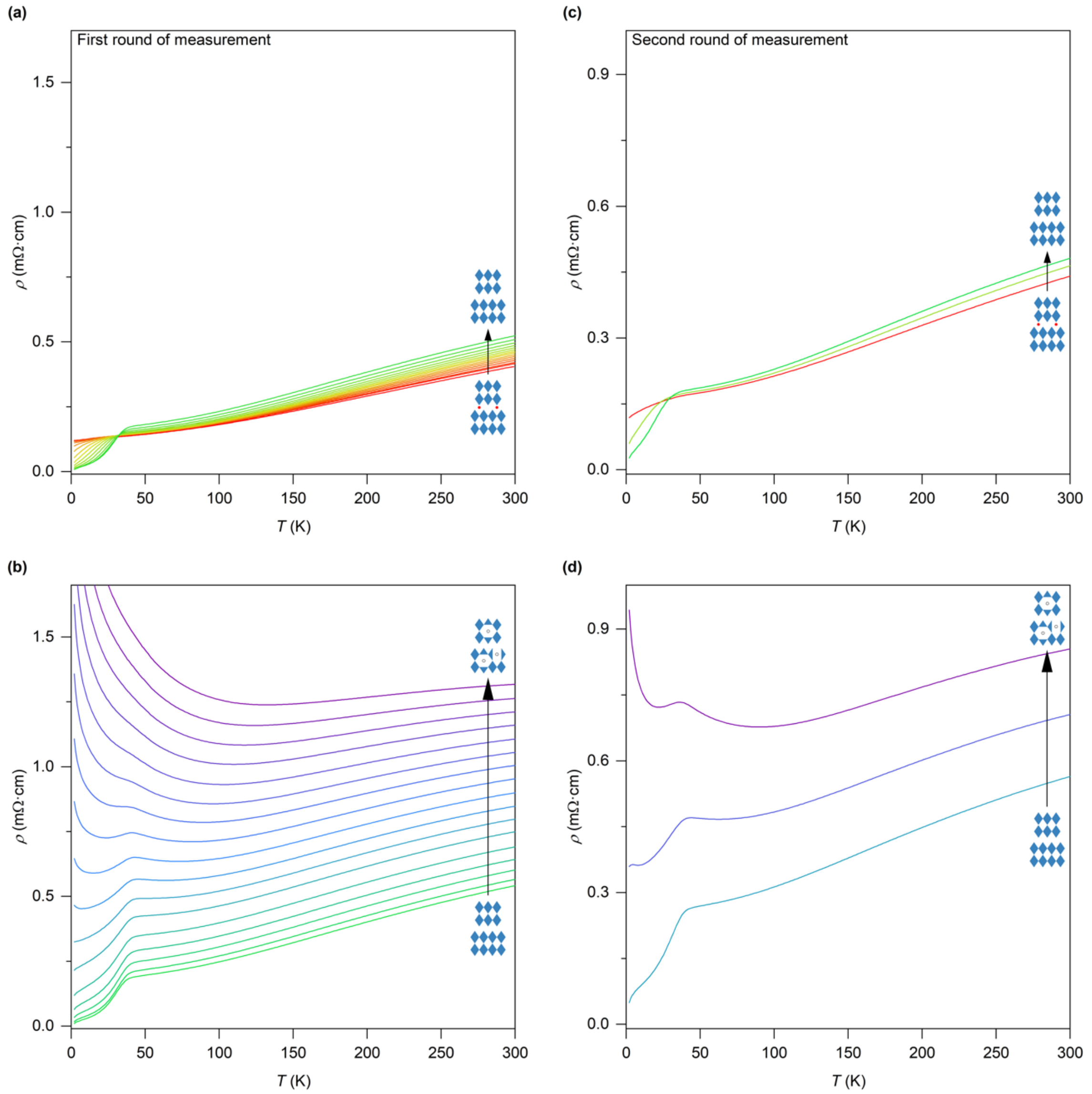

FIG. S4. Reversibility of continuous oxygen tuning. (a)-(d) $\rho(T)$ of LPNO thin films subjected to a first and second cycle of ozone annealing, measured after sequential vacuum annealing. (a), (c) show the evolution from the metallic state to the superconducting state (Stage I to 'Crossover'), while (b), (d) show further evolution from the superconducting state to the insulating state ('Crossover' to Stage II). (a)-(b) correspond to the first-round ozone-annealed sample while (c)-(d) represent the second-round ozone-annealed sample. The inset schematics illustrate the structural response to changes in oxygen stoichiometry. Although the crystallinity somewhat degraded after the second ozone annealing, a similar evolution in transport properties was observed in both measurement rounds.

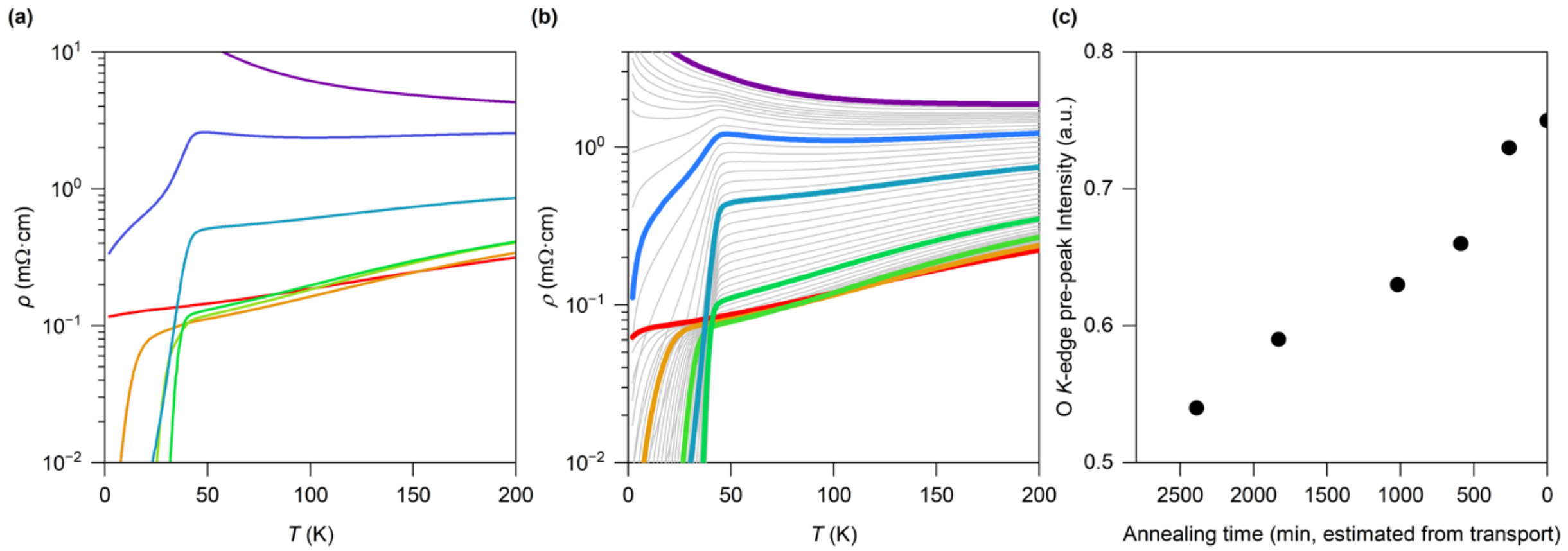

FIG. S5. Calibration of sample-dependent annealing time with spectroscopic quantities reflecting oxygen stoichiometry. (a) $\rho(T)$ curves of the LSNO films in Fig. 3(b),(c), with colors denoting different oxygen stoichiometries. (b) $\rho(T)$ curves replotted from Fig. 4, with thick, color-coded curves denoting the transport curves that are best matched to those in (a). (c) O $K$-edge pre-peak intensity of the samples in (a) versus the annealing time of the corresponding matched curve in (b).

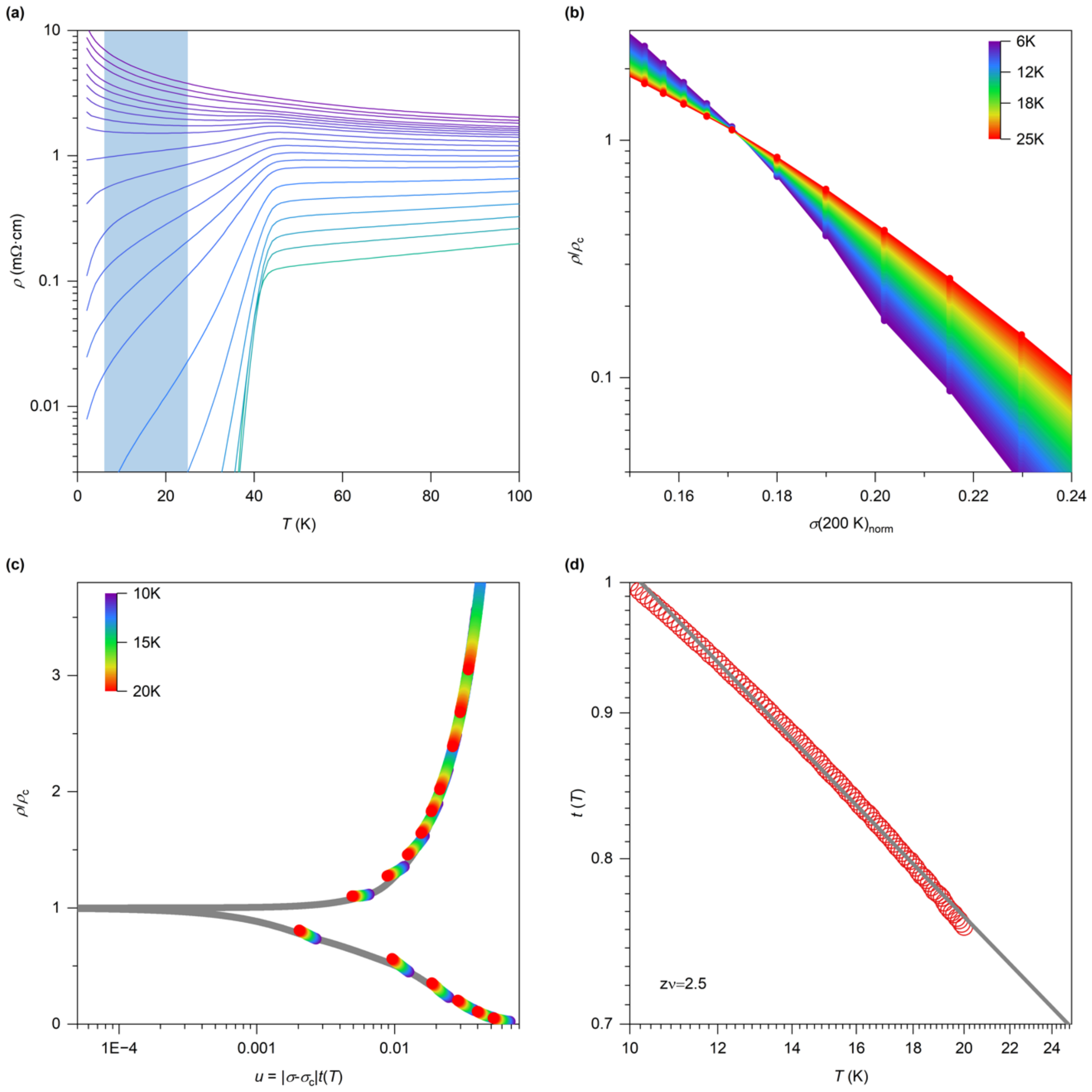

FIG. S6. Scaling analysis of the SIT in bilayer nickelate thin films. (a) $\rho(T)$ of ozone-annealed LSNO thin films measured after sequential vacuum annealing in Stage II (the same sample shown in Fig. 4). The blue shaded region shows the temperature range for the finite-size scaling analysis. (b) The same dataset in (a) plotted as a function of $\sigma(200\ \mathrm{K})_{\mathrm{norm}}$ at fixed temperatures between 6 K and 25 K, with each color representing a fixed temperature. Continuous curves are interpolations of data points at different temperatures. All curves cross at the critical resistance $\rho_c = 13.8$ kΩ/bilayer. (c) Scaling of the same data with respect to variable $u = |\sigma - \sigma_c| t(T)$. A single set of temperature-dependent parameters $t(T)$ can collapse all data to a universal scaling function on both sides of the SIT. (d) $t(T)$ obtained from finite-size scaling analysis in (c). The slope of the line fit (grey line) yields the critical exponents product $z\nu = 2.5$.

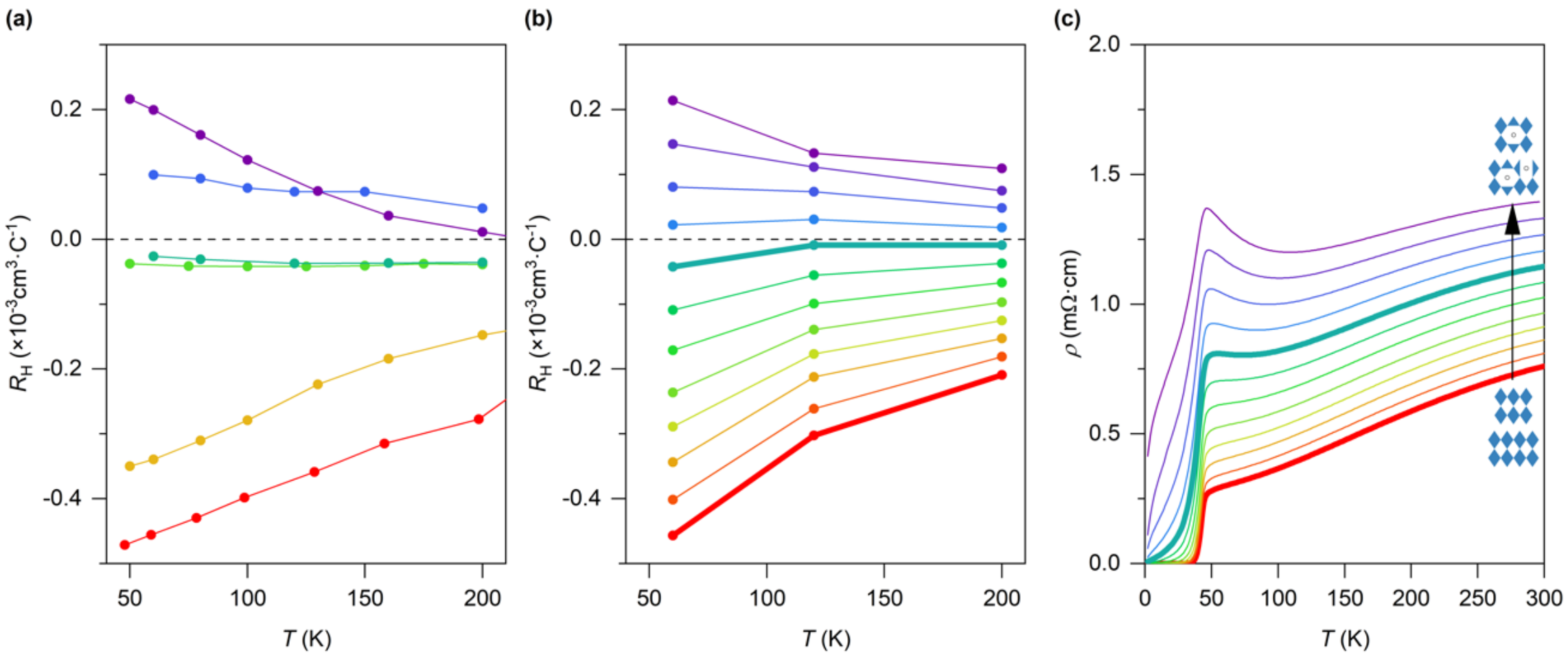

FIG. S7. Evolution of the Hall coefficient with temperature ($R_H(T)$) in bilayer nickelate thin films with different oxygen stoichiometries. (a) $R_H(T)$ data from refs. [6,45,7,21,23,20] (ordered top to bottom). (b) $R_H(T)$ for the sample shown in Fig. 4, tuned across oxygen stoichiometry near the stoichiometric composition; colors denote different oxygen stoichiometries. (c) $\rho(T)$ of the same sample as in (b) at various oxygen stoichiometries. Curves corresponding to the same stoichiometry share the same color. The schematic on the right illustrates the structural changes associated with varying oxygen stoichiometry. The thick red curves in (b) and (c) correspond to the optimally superconducting state, whereas the thick green curves represent a slightly oxygen-deficient state whose near-zero $R_H(T)$ more closely resembles that reported in [7,21].

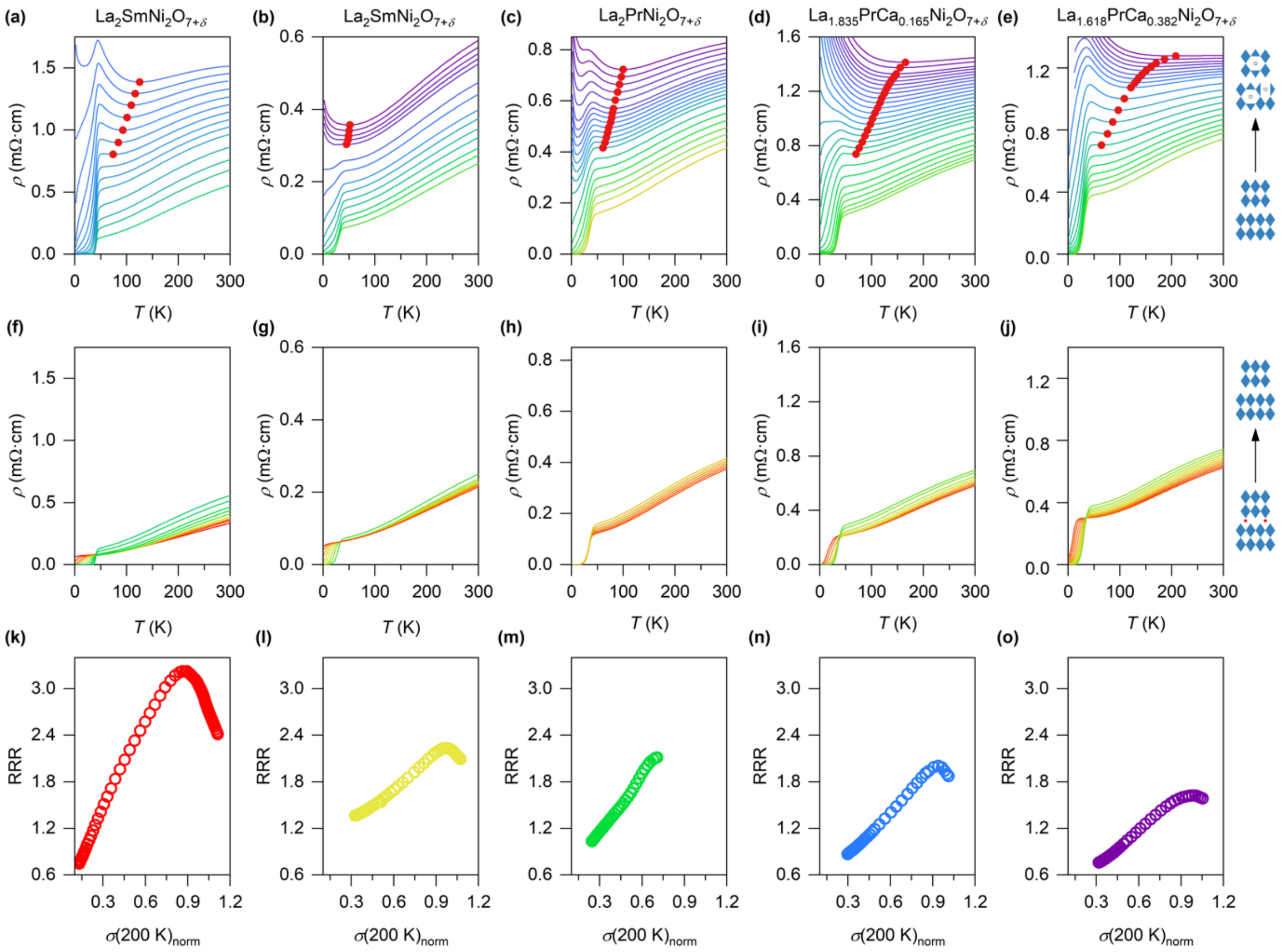

FIG. S8. Evolution of transport properties in bilayer nickelate thin films with different oxygen stoichiometries. Each column corresponds to one of the five samples in Fig. 1(a), identified above panels (a)-(e). (a)-(j) Selected $\rho(T)$ curves, with colors denoting different oxygen stoichiometries. (f)-(j) show the evolution from the metallic state to optimal superconductivity (Stage I to 'Crossover'). (a)-(e) show further evolution from the optimally superconducting state to the insulating state ('Crossover' to Stage II). $T_{\mathrm{ins}}$, which is sensitive to oxygen deficiency [54], is indicated by filled red circles. The schematics on the right illustrate the structural evolution driven by changes in oxygen stoichiometry. (k)-(o) Evolution of RRR plotted against $\sigma(200\text{ K})_{\mathrm{norm}}$ for the sample in the corresponding column, color coded as in Fig. 1(a).